\documentclass[a4paper,11pt]{article}
\usepackage{pos}
\usepackage{booktabs}
\usepackage{caption}
\usepackage{subcaption}
\usepackage{xcolor}

\title{Sterile neutrino dark matter in the super-weak model}
\ShortTitle{Sterile neutrino dark matter in SWSM}

\author*[a]{Károly Seller}
\affiliation[a]{ELTE Eötvös Loránd University,\\
  Pázmány Péter sétány 1/A, H-1117 Budapest, Hungary}

\emailAdd{karoly.seller@ttk.elte.hu}
\abstract{The {\it super-weak model} is a U(1) extension of the Standard Model.
In addition to a mediator $Z'$, a singlet scalar field $\chi$ is added to deal with the meta-stability of the SM vacuum, and right-handed neutrinos are introduced to account for the non-vanishing neutrino masses. 
We demonstrate that the lightest right-handed neutrino is a possible dark matter candidate with masses of $\mathcal{O}(10)~\mathrm{MeV}$ via the resonant freeze-out scenario. 
The dark matter production is explored and the available parameter space is discussed from the point of view of various experiments and beyond SM aspects.}

\FullConference{%
  41st International Conference on High Energy physics - ICHEP2022\\
  6-13 July, 2022\\
  Bologna, Italy
}

\begin{document}
\maketitle

\section{Introduction}

Despite the overwhelming theoretical and experimental success of the Standard Model (SM), there are multiple high-significance observations which are not explained within its framework.
One such observation that calls for an extension to the SM is the existence of a seemingly invisible matter that permeates the Universe, which we call dark matter.
The currently accepted cosmological model as well as astrophysical observations predict that dark matter is highly abundant, providing roughly 5 times the energy density of regular matter.
As the existence of dark matter is an experimentally well established fact, the search for models to theoretically describe the nature of dark matter has been a highly active field in physics.
One has to keep in mind however that dark matter is only one side of the beyond SM (BSM) puzzle.
As such, we have to make sure that any modification we make to the SM does not contradict other observations.
In particular, an interesting direction to take is to find what the minimal extension to the SM that is 
capable of explaining multiple (all if possible) observed phenomena.
The {\it super-weak} extension of the standard model (SWSM) \cite{Trocsanyi:2018bkm} is an economical extension to the SM which aims to achieve this goal.

\section{The super-weak model}

The super-weak model is a U(1)$_z$ extension of the SM originally introduced in Ref.~\cite{Trocsanyi:2018bkm}.
The particle spectrum involves the new gauge boson $Z'$, 3 right-handed neutrinos $N_i$, and a singlet scalar $\chi$, in additional to the degrees of freedom of the SM.
All particles within the spectrum are charged under the new U(1)$_z$ symmetry.

The model is designed to be a simple renormalizable extension capable of explaining multiple BSM observations.
Neutrino masses and oscillation have been studied in Refs.~\cite{Iwamoto:2021wko,Karkkainen:2021syu}, vacuum stability and cosmic inflation was investigated in Refs.~\cite{Peli:2019vtp,Peli:2022ybi}, and dark matter production was discussed in Refs.~\cite{Iwamoto:2021fup,Seller:2021csy}. 
The possibility of leptogenesis in the super-weak model is currently being explored, along with a global fit of all super-weak parameters to experimental results.

The concept behind the super-weak model is to explore whether a simple extension with a few parameters can provide a non-vanishing parameter space when fit to experimental results.
However, one should realize that both positive and negative answers to this question are exciting.
If the model is capable of dealing with the currently known BSM observations then it shows that simplicity can still be a good guideline, as it used to be historically in particle physics.
On the other hand, if the answer is negative then we have multiple exciting possibilities to think about:
(i) either the questions we asked are not all related to particle physics, and thus cannot be solved simultaneously by just an extension of the SM, or
(ii) a more complex extension is required with more parameters and particles predicting exciting but highly complex phenomenology, or
(iii) we are facing some fundamental problem in the theory.

In the following subsections we describe the interactions of the new particles and the parameters of the model.

\subsection{Gauge sector}

In the super-weak model the SM gauge group is extended by a U(1)$_z$ symmetry group.
Via spontaneous symmetry breaking the full symmetry group is broken to the one observed today, SU(3)$_\mathrm{c}\otimes$U(1)$_\mathrm{em}$.
This symmetry breaking is expected to happen in 2 steps: first the U(1)$_z$ symmetry is broken by the non-zero vacuum expectation value of the singlet scalar $\chi$, then the Higgs breaks the electroweak sector down to U(1)$_\mathrm{em}$ as in the SM.

After spontaneous symmetry breaking the gauge eigenstates ($B_\mu$, $W^3_\mu$, $B'_\mu$) are rotated to the mass eigenstates ($A_\mu$, $Z_\mu$, $Z'_\mu$).
The rotation is characterized by two angles (the third one is unphysical): the Weinberg angle $\theta_\mathrm{W}$ that rotates on the $(B_\mu,~W^3_\mu)$ plane, and $\theta_Z$ that rotates on the $(W^3_\mu,~B'_\mu)$ plane.

%\begin{table}[t]
%    \centering
%    \begin{tabular}{cccc}\toprule
%         & SU(2)$_\mathrm{L}$ & U(1)$_y$ & U(1)$_z$ \\
%        \midrule
%        $Q_\mathrm{L}$ & $\mathbf{2}$ & $1/6$ & $1/6$ \\
%        $U_\mathrm{R}$ & $\mathbf{1}$ & $2/3$ & $7/6$ \\
%        $D_\mathrm{R}$ & $\mathbf{1}$ & $-1/3$ & $-5/6$ \\
%        $L_\mathrm{L}$ & $\mathbf{2}$ & $-1/2$ & $-1/2$ \\
%        $N_\mathrm{R}$ & $\mathbf{1}$ & 0 & $1/2$ \\
%        $e_\mathrm{R}$ & $\mathbf{1}$ & $-1$ & $-3/2$ \\
%        \midrule
%        $\phi$ & $\mathbf{2}$ & $1/2$ & 1 \\
%        $\chi$ & $\mathbf{1}$ & 0 & $-1$ \\
%        \bottomrule
%    \end{tabular}
%    \caption{Particle content and charge assignment of the super-weak model.}
%    \label{tab:ChargeAssignment}
%\end{table}

The gauge interactions are given by the covariant derivative in conjunction with the charges defined in Table~1 of Ref.~\cite{Trocsanyi:2018bkm}.
For a fermion $\psi_f$ the neutral sector of the gauge covariant derivative is
\begin{equation}
    \mathcal{D}^\mathrm{(neutral)}_\mu\psi_f=\Big(\partial_\mu + i\mathcal{Q}_f^{(A)}A_\mu + i\mathcal{Q}^{(Z)}_fZ_\mu + i\mathcal{Q}_f^{(Z')}Z'_\mu\Big)\psi_f\,,
\end{equation}
where the charges are given as
\begin{subequations}
\begin{equation}
    \mathcal{Q}_f^{(A)}=(T^3_f+y_f)|e|
\end{equation}
\begin{equation}
    \label{eq:neutralcurrent}
    \mathcal{Q}_f^{(Z)}=(T^3_f\cos^2\theta_\mathrm{W}-y_f\sin^2\theta_\mathrm{W})g_{Z^0}\cos\theta_Z-z_fg_z\sin\theta_z
\end{equation}
\begin{equation}
    \mathcal{Q}_f^{(Z')}=(T^3_f\cos^2\theta_\mathrm{W}-y_f\sin^2\theta_\mathrm{W})g_{Z^0}\sin\theta_Z+z_fg_z\cos\theta_z\,.
\end{equation}
\end{subequations}
Here $y_f$ and $z_f$ is the hypercharge (the eigenvalue of $Y/2$) and super-weak charge, $T^3_f$ is the weak isospin for the fermion $\psi_f$, and $g_{Z^0}=\sqrt{g^2+g'^2}$ is the weak coupling.

The gauge sector seemingly has 2 new parameters, the U(1)$_z$ gauge coupling $g_z$, and the mixing angle $\theta_Z$.
However they are not independent, it can be shown that
\begin{equation}
    \tan(2\theta_Z)=\frac{4z_\phi g_z}{g_{Z^0}}+\mathcal{O}\left(\frac{g_z^2}{g_{Z^0}^2}\right)\,.
\end{equation}
From the smallness of the super-weak coupling the smallness of the mixing angle follows, thus the new interaction only modifies the weak neutral current (c.f. Eq.~\eqref{eq:neutralcurrent}) at a level of $\mathcal{O}\left(g_z^2/g_{Z^0}^2\right)$, which cannot be observed for sufficiently small couplings.

\section{Dark matter production}

In the super-weak model the lightest sterile neutrino $N_1$ is the dark matter candidate.
The correct dark matter abundance is reached primarily through the so-called vector boson portal via $Z'$ decays.
We should note however that this is not the only way: (i) due to the extra scalar mixing with the SM Higgs field the \emph{scalar portal} is also available, but suppressed as compared to the vector portal; (ii) the model has sterile neutrinos, thus the \emph{neutrino portal} is also available.
We shall only focus on the main production channels via $Z'$ exchange.

Dark matter abundance can be reached in various manners depending on the production mechanism.
Two such scenarios are the freeze-out and freeze-in mechanism.
In the former, dark matter reaches equilibrium with the cosmic plasma at some high temperature due to rapid interactions between the dark matter particles and the SM.
At some low temperature $T\simeq m_\mathrm{DM}/20$ the interactions become slow compared to the expansion of the Universe and dark matter freezes out.
In the freeze-in case, dark matter never reaches chemical equilibrium with the SM due to feeble couplings. In this case the main production channels are decays into dark matter, which terminate (dark matter freezes in) when the decay sources vanish.
We will now focus on the freeze-out scenario and refer to Ref.~\cite{Iwamoto:2021fup} for the freeze-in study.

In the freeze-out scenario scatterings play the major role in creating the required dark matter abundance.
The super-weak model provides gauge interactions between the SM fermions and the dark matter candidate $N_1$ as described in the previous section.
We assume a light sterile neutrino with mass of $\mathcal{O}(10)~$MeV. 
It follows that the relevant interactions in terms of dark matter are those with electrons and SM neutrinos.

Dark matter is generally overproduced in the freeze-out scenarios, thus in the super-weak model we employ resonant production to evade stringent constraints on the coupling.
Resonant production exploits the resonance peak in the $s$-channel cross sections: the peak width is naturally small due to the $Z'$ decay rate being proportional to $g_z^2$.
To find the thermal rate we have to integrate over the cross section in the physical region of $s$, in particular in the Maxwell-Boltzmann approximation
\begin{equation}
    \gamma(\psi_f\psi_f\to N_1N_1)=\frac{T}{64\pi^2}\int_{s_\mathrm{min}}^\infty\mathrm{d}s~\hat{\sigma}(s)\sqrt{s}K_1\left(\frac{\sqrt{s}}{T}\right)\,.
\end{equation}
Here $\hat{\sigma}(s)=2\lambda(s,m_f^2,m_f^2)\sigma(s)/s$ is the reduced cross section, $\lambda(a,b,c)$ is the Källén-function, and $K_1(x)$ is the modified Bessel function of the second kind.
Since $\sigma(s)$ is highly peaked around $\sqrt{s}=M_{Z'}$ it can dominate the integral, which we call \emph{resonant production}.
We have two requirements for resonant production: (i) we need the peak to be in the physical region, i.e., $M_{Z'}^2\geq s_\mathrm{min}=4m_{N_1}^2$, and (ii) we need the argument of the Bessel function to be small at decoupling, i.e. $\left.\sqrt{s}/T\right|_{s=M_{Z'}^2,T=T_\mathrm{dec}}=\mathcal{O}(1)$.
The second point is required because $K_1(x)$ is exponentially small for large arguments, thus it can suppress the peak contribution in the integral for large $x$.
These points provide a preferred relation between the mediator and dark matter masses, in particular for resonance domination we need $M_{Z'}\approx 2m_{N_1}$.

\begin{figure}
    \centering
    \begin{subfigure}[b]{0.48\textwidth}
        \includegraphics[width=\textwidth]{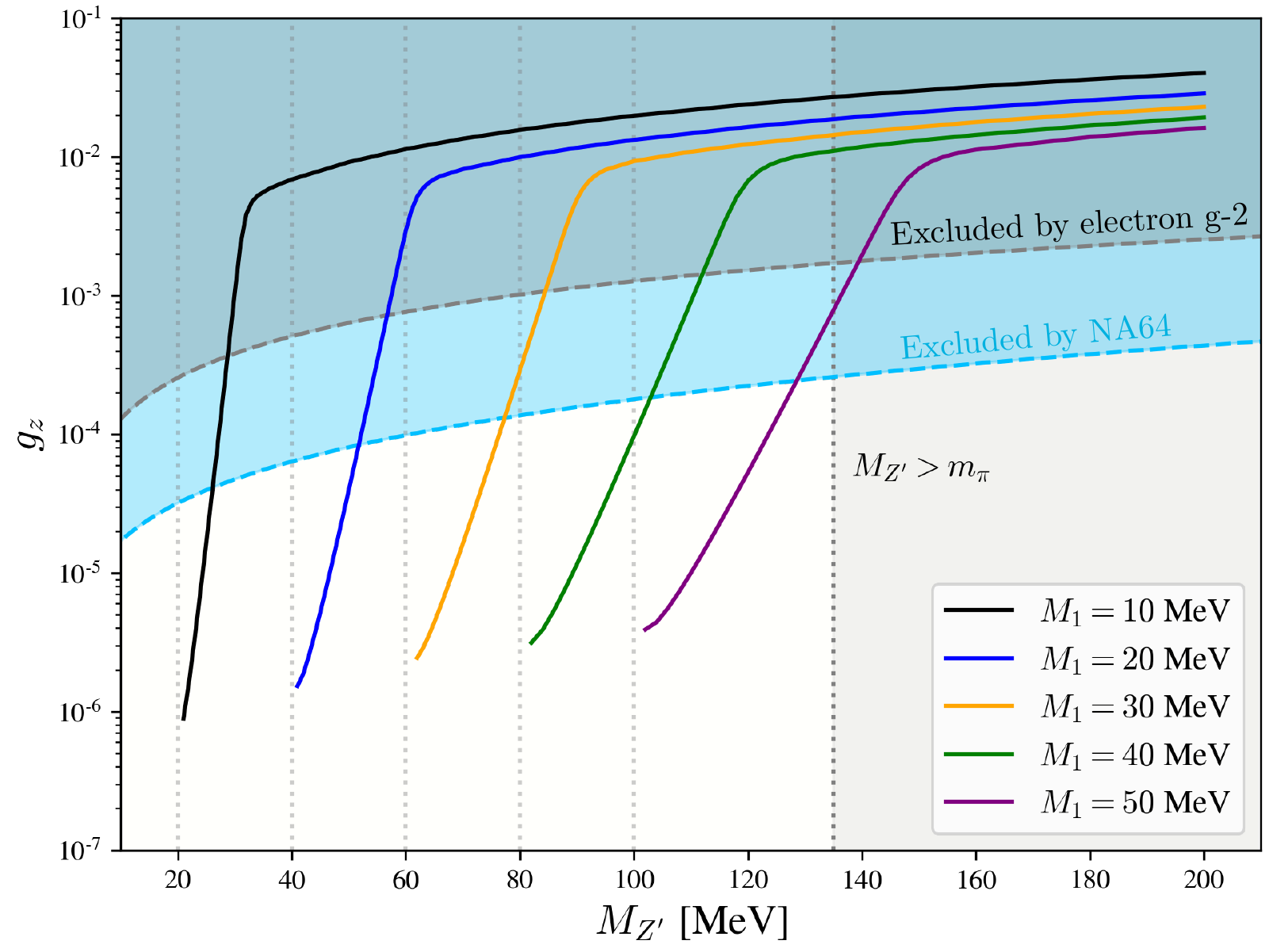}
    \end{subfigure}
    ~
    \begin{subfigure}[b]{0.46\textwidth}
        \includegraphics[width=\textwidth]{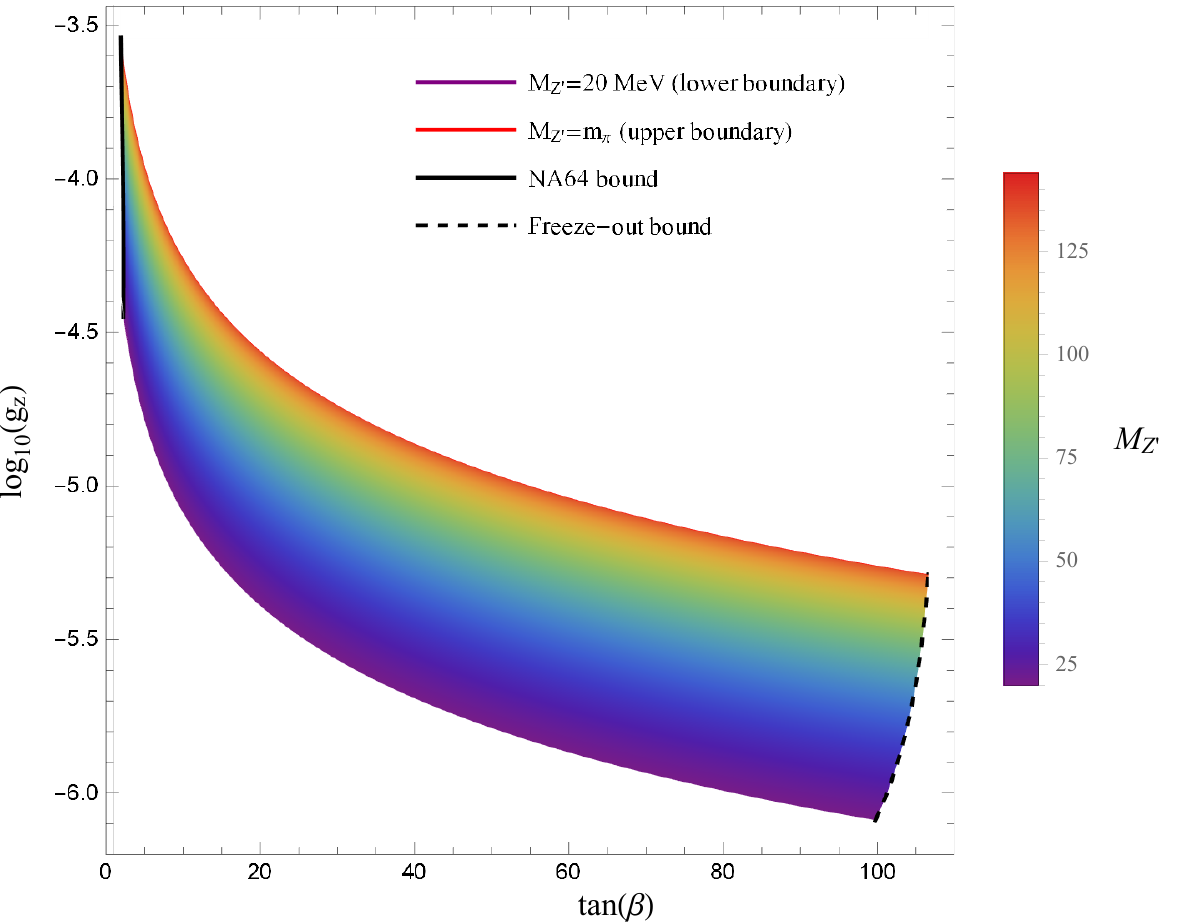}
    \end{subfigure}
    \caption{\emph{Left panel:} The parameter space of the freeze-out production of sterile neutrino dark matter in the super-weak model. The solid lines indicate the required couplings $g_z$ in terms of the U(1)$_z$ gauge boson mass $M_{Z'}$ for various values of the dark matter mass $M_1$. \emph{Right panel:} The favoured parameter space shown in the left panel translated to the $g_z$ versus $\tan\beta=w/v$ plane, where $w$ is the vacuum expectation value of the scalar singlet. We have used $M_{Z'}\in [20~\mathrm{MeV},m_\pi]$, which provide the upper and lower boundaries in the figure, whereas small values of $\tan\beta$ are ruled out by NA64, and large values of $\tan\beta$ are disfavored by the overproduction of dark matter.}
    \label{fig:ParameterSpace}
\end{figure}

We solve the Boltzmann equations with 3 free parameters: (i) the super-weak coupling $g_z$, (ii) the mass of the dark matter candidate sterile neutrino $m_{N_1}\equiv M_1$, and (iii) the mass of the new gauge boson $M_{Z'}$.
The resulting parameter triplets reproducing the measured dark matter energy density ($\Omega_\mathrm{DM}=0.265$) are shown in the left panel of Fig.~\ref{fig:ParameterSpace}.
The effects of resonant amplification of the thermal rate are clearly visible in the region where $M_{Z'}\approx 2M_1$ (steep sections of solid lines).
Resonant effects provide a viable parameter region for which dark matter may be fully constituted by the lightest sterile neutrino.
Note that without resonant effects the required coupling is $g_z=\mathcal{O}(10^{-1}-10^{-2})$ which is excluded by multiple well established experimental constraints.

\section{Experimental constraints}

Constraints on the model can be roughly categorized into two groups: constraints from particle physics experiments and from astrophysical observations.
Particle physics experiments put upper bounds on the effects some new particle can have through precision measurements of a certain observable, or non-observation of BSM phenomena.
Astrophysical observations are usually less stringent due to the large theoretical and experimental uncertainties of these measurements.
Nevertheless, they are still good guidelines for a model to follow.
Here we consider two constraints, one from the particle physics side, and one from the astrophysical side, for more details see Refs.~\cite{Iwamoto:2021fup,Seller:2021csy}.

The most important constraint on the freeze-out dark matter production in the super-weak model comes from the NA64 experiment \cite{Banerjee:2019pds}.
Their bounds are due to the non-observation of dark photon producing Bremsstrahlung processes.
In practice, they look for missing energy events in an experimental setup where an electron beam is fired at a fixed target.
With the non-observation of significant missing energy, the NA64 group produces a bound on the kinetic mixing between the photon and the dark photon.
This constraint is then translated to the super-weak model as a bound $g_z^{\mathrm{max}}(M_{Z'})$, the resulting bound is shown in the left panel of Fig.~\ref{fig:ParameterSpace}.

From the astrophysical side, the success of the SM Big Bang nucleosynthesis (BBN) provides us with constraints any new particle can have around $T=1~$MeV.
In principle, new particles can modify either the Hubble expansion through introducing extra degrees of freedom in the Universe, or the particle interactions.
The former is easily avoided since the only beyond the SM particles present at BBN are the $N_1$ neutrinos, but their abundance is many orders of magnitude smaller than those in equilibrium. 
The latter is generally more complicated due to the many possible interactions in the BBN.
The super-weak model does not modify the charged weak current, thus BBN processes are safe. 
However, if the $Z'$ is too heavy, then it may decay into pions: an overabundance of pions will severely affect the proton-to-neutron ratio which is an essential parameter for BBN.
To avoid completely such a breakdown of BBN, we have put an upper limit on the $Z'$ mass at $M_{Z'}=m_\pi$.

\section{Outlook}

As described in the introduction, the super-weak model is designed to solve multiple beyond the SM problems with a single, economic extension of the SM.
As for cosmology, after the successful incorporation of dark matter production in the model, we can explore the possibility of leptogenesis.
While this problem is somewhat decoupled from the dark matter, we have to keep in mind that we are thinking within the context of a single model, thus the favored couplings and masses we have found from the DM searches should apply to the leptogenesis scenarios as well.
In particular, the $Z'$ obtains its mass through the spontaneous breaking of the U(1)$_z$ symmetry, where the vacuum expectation value $w$ of the singlet scalar is an essential parameter for leptogenesis (it sets the scale).
Dark matter searches indicate that not all $Z'$ masses are allowed, thus the allowed parameter space on $g_z(M_{Z'})$ can be translated to $g_z(w)$.
The resulting region is shown in the right panel of Fig.~\ref{fig:ParameterSpace}.
In the figure, $\tan\beta=w/v$ is allowed to lie within the range $2<\tan\beta<100$, setting the scale of leptogenesis to be relatively low.

\section{Conclusion}

We have shown that the super-weak model can provide a viable parameter space for sterile neutrino dark matter.
This parameter region is going to be tested by particle physics experiments such as LHCb \cite{Ilten:2015hya} within the next few years.
A detailed analysis of whether this parameter space is consistent with leptogenesis is an ongoing project.

\end{document}